\documentclass{intech07}

\booktitle{Will-be-set-by-IN-TECH}%

\chaptertitle{Experimental Study of the Intrinsic and Extrinsic Transport Properties of Graphite and Multigraphene Samples} 

\authors{J. Barzola-Quiquia, A. Ballestar,  S. Dusari  and P. Esquinazi}
\affiliation{Division of Superconductivity and Magnetism,
Institute for Experimental Physics II, University of Leipzig}
\country{Germany }

\begin{document}

\maketitle

\section{Introduction}

This chapter\footnote{This work is supported by the DFG under DFG
ES 86/16-1. A.B. is supported by the ESF "Nachwuchsforschergruppe"
"Funktionale multiskalige Strukturen" and S.D. by the Graduate
school of Natural Sciences of the University of Leipzig
"BuildMona".} deals with the following basic questions, which
sometimes surprise or even irritate the reader, namely: Which are
the intrinsic properties of the graphene layers inside the
graphite structure? Are their transport properties quasi-two
dimensional and are they comparable with those of single layer
graphene? Which is the coupling between ideal, i.e. defect free,
graphene layers inside graphite? Which is the influence of defects
and interfaces within the graphene and graphite structure? These
basic questions remain in the literature still unanswered to a
large extent mainly because the influence on the transport
properties of defects and sample or crystal sizes was not
systematically taken into account.

A large amount of the published interpretations on the
experimental transport data of real graphite samples relied on the
assumption that these were intrinsic properties. Let us start with
a rather simple example. In the last fifty years scientists
flooded the literature with reports on different kinds of
electronic measurements on graphite samples, providing evidence
for carrier (electron plus hole) densities {\it per graphene
layer} in graphite\footnote{In this chapter and to facilitate a
direct comparison between the carrier density of a single graphene
layer and that one inside the graphite structure, the carrier
density per area will be used. This can be easily obtained
multiplying $n$, the three dimensional carrier density, by the
distance between graphene layers in graphite $a_0 = 0.335~$nm.
This estimate assume that the coupling between graphene layers
inside graphite is very weak and that the electron transport is
mainly on the two dimensional graphene layers.} and at low
temperatures $n_0
> 10^{10}~$cm$^{-2}$ as one can read in the standard book from
\cite{kelly} or the old publication from \cite{mcc64}, or from
more recent work by \cite{gru08} ($n_0 \simeq 10^{12}~$cm$^{-2}$)
or \cite{kum10} ($n_0 \simeq 2.4 \times 10^{11}~$cm$^{-2}$). But
why this concentration is not a constant and does apparently
depend on the measured sample? The reader may ask then why at all
is the knowledge of the intrinsic carrier concentration $n_0$ so
important?

The carrier concentration is one of those basic parameters needed
to estimate several others necessary to get a reliable picture of
the electronic band structure and from this to understand all
intrinsic (or not) transport properties. For example,
two-dimensional (2D) calculations assuming a coupling $\gamma_0$
between nearest in-plane neighbors C-atoms on the graphene plane
give a carrier density (per C-atom) $n(T) = (0.3 \ldots 0.4) (k_B
T/\gamma_0)^2$ ($\gamma_0 \simeq 3~$eV and $T$ is the temperature)
\citep{kelly}. Because all experimental values obtained from bulk
graphite samples indicated  a finite $n(T \rightarrow 0) = n_0
> 0$ then the straightforward and "easiest" solution to solve
this "disagreement" is to start including more free parameters in
the tight-binding electronic band structure calculations.

For example, introducing a new coupling $\gamma_1$ between C-atoms
of type $\alpha$ in adjacent planes one obtains $n(T) = a
(\gamma_1/\gamma_0^2) T + b (T/\gamma_0)^2 + c
(T^3/\gamma_0^2\gamma_1) + \ldots$ ($a,b,c,\ldots$ are numerical
constants), where the "accepted" value for $\gamma_1 \sim 0.3~$eV.
Also in this case $n(T \rightarrow 0) \rightarrow 0$. We stress
that neither in single layer graphene nor in  graphite such
$T-$dependence was ever reported \footnote{It is interesting to
note that the carrier concentration obtained in bulk graphite by
\protect\cite{gar08}, using an original and parameter-free method
to determine it and the mean free path, can be fitted up to $\sim
200~$K by $n[$cm$^{-2}] \simeq
    n_0  + 10^5 T^2 + 7.5 \times 10^3 T^3$ with $T$ in [K]
    and $n_0 \simeq 2 \times 10^8~$cm$^{-2}$. The same data, however,
    can be also well explained by
    a semiconducting-like exponential function $\exp(-E_g/2T)$ with an energy gap $E_g \sim 50~$meV.},
    i.e. a large density background $n_0$ was
always measured and assumed as "intrinsic" without taking care of
any influence from lattice defects or impurities. To fit
experimental data and obtain a finite Fermi energy $E_F$ -- in the
simplest case $E_F \sim \gamma_2$ \citep{kelly, dil77} -- up to
seven free parameters were and still are introduced, even for
carrier density as small as $n \simeq |-8 \times 10^9|~$cm$^{-2}
(E_F \simeq -29~$meV) as obtained recently from magnetotransport
data in bulk pyrolytic graphite \citep{sch09}.

Taking into account the exhaustive experience accumulated in
gapless or narrow gap semiconductors \citep{tsi97} we should
actually expect that the measured carrier density $n_0\lesssim
10^{12}~$cm$^{-2}$ is not intrinsic of the graphite structure but
it is influenced by impurities and/or defects in the
graphite/graphene lattice. The reader should keep in mind that a
carrier density of the order of $10^8$~cm$^{-2}$ means one carrier
in $1~\mu$m$^2$ graphene area, which could be produced actually by
having a single vacancy or impurity in the same graphene area, in
case one carrier is generated by one of these defects, as
experimental \citep{arn09} and theoretical \citep{sta07} work
suggests. Experimental evidence published recently and partially
reviewed in this chapter speaks against an intrinsic origin of -
even a part of - the measured $n_0$  in graphite samples,  casting
doubts on the relevance of related electronic band structure
parameters obtained in the past. On the other hand this new
knowledge will help significantly to clarify observed transport
phenomena.

In Section~\ref{ballistic} of this chapter we describe a method
that one can use to obtain the carrier density,
 the mean free path and the mobility of the carriers inside graphite
 without free parameters. In that Section we review a systematic study of the
transport in small multigraphene\footnote{We use this word to
refer to graphite samples of not more than a few micrometers in
length and width and thickness below $\sim 100$~nm. The reason for
this kind of geometrical restriction will become clear in
Section~\ref{sample}.}  samples that reveals room-temperature
mobility values ($\sim 6 \times 10^7~$cm$^2$V$^{-1}$s$^{-1}$) per
graphene layer inside graphite, which overwhelm those reported in
literature for single graphene layers, indicating the higher
quality of the graphene layers inside ideal graphite. This quality
is also reflected by the extremely low room-temperature carrier
density ($\sim 7 \times 10^8~$cm$^{-2}$) obtained for good, but
certainly not ideal, quality multigraphene samples. These studies
indicate that ballistic transport with mean free path in the
micrometer range is possible in graphite at room temperature.

In Section \ref{sample} we describe the main characteristics  of
bulk and multigraphene samples, their characterization using
transmission electron microscopy (TEM), electron backscattering
diffraction (EBSD), electronic transport and Raman spectroscopy.
We show the correlations between the internal microstructure and
sample size -- lateral as well as thickness from millimeter size
graphite samples to  mesoscopic ones, i.e. tens of nanometer thick
multigraphene samples -- and the temperature ($T$) and magnetic
field ($B$) dependence of the longitudinal resistivity
$\rho(T,B)$.  Low energy transmission electron microscopy reveals
that the original highly oriented pyrolytic graphite (HOPG)
material - from which the multigraphene samples were obtained by
exfoliation - is composed of a stack of $\sim 50$~nm thick and
micrometer long crystalline regions separated by interfaces
running parallel to the graphene planes \citep{bar08}. We found a
qualitative and quantitative change in the behavior of $\rho(T,B)$
 upon thickness of the samples,
indicating that their internal microstructure is important. The
overall results described in sections~\ref{sample} and \ref{super}
indicate that the metallic-like behavior of $\rho(T)$ at zero
magnetic field measured for bulk graphite samples is not intrinsic
of ideal graphite.

The influence of internal interfaces on the transport properties
of bulk graphite is described in detail in Section~\ref{super} of
this chapter. We will show that in specially prepared
multigraphene samples the transport properties show clear signs
for the existence of granular superconductivity within the
graphite interfaces, which existence was firstly reported
by~\cite{bar08}. Based on the results described in
Section~\ref{super} we argue that the superconducting-insulator or
metal-insulator transition (MIT) reported in literature for bulk
graphite is not intrinsic of the graphite structure but it is due
to the influence of these interfaces.

\section{Samples characteristics}\label{sample}

\subsection{Sample preparation}

In order to systematically study the transport properties of ideal
graphite and compare them with those of bulk graphite samples
measured in the past, we need to perform measurements in different
tens of nanometer thick multigraphene samples of several
micrometer square area. The samples presented in this chapter were
obtained from a highly oriented pyrolytic graphite (HOPG) bulk
sample with a mosaicity (rocking curve width) of $\sim
0.35^{\circ} \pm 0.1^{\circ}$ from Advanced Ceramics company. This
material does not only guaranty high crystalline quality but also
allows us to easily cleave it and obtain up to several hundreds of
micrometers large flakes  with thickness from a few to several
tens of nanometers. The starting geometry of the bulk graphite
material for the preparation of the flakes was $\sim 1~$mm$^2$ and
$\sim 20~\mu$m thickness. The selected piece was located between
two substrates and carefully pressed and rubbed. As substrate we
used p-doped Si with a 150~nm SiN layer on top. The usefulness of
the SiN layer on the Si substrate is twofold: firstly the
multigraphene flake on it shines with high contrast by
illuminating it with white light allowing us to use optical
microscopy to select the multigraphene samples.

\begin{figure}[t] 
\centering
\includegraphics[width=8.5cm]{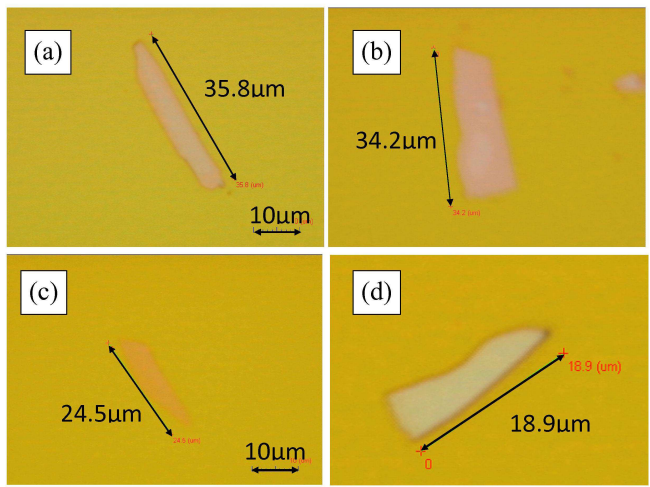} 
\caption{(a) to (d): Optical microscope pictures of four
multigraphene samples prepared as described in the text. The
dimensions of the samples can be seen directly at the pictures.
The thickness of the samples are: (a) 60~nm, (b) 55~nm, (c) 10~nm,
and (d) 85~nm.}\label{figsamples}
\end{figure}

Immediately after the rubbing process we put the substrate
containing the multigraphene films on it in a ultrasonic bath
during 2~min using high concentrated acetone. This process cleans
and helps to select only the good adhered multigraphene films on
the substrate. After this process we used optical microscopy and
later scanning electron microscopy (SEM) to select and mark the
position of the films. Figure~\ref{figsamples} shows four of the
investigated samples of different micrometer length and tens of
nanometer thickness.

For the preparation of the electrical contacts we used
conventional electron beam lithography. The contacts were done by
thermal deposition of Pd (99.95\%) or Pt/Au (99.5\%/99.99\%) in
high vacuum conditions. We have used Pd or Pt/Au because these
elements do not show any significant Schottky barrier when used
with graphite. This has been checked by $I-V$ measurements in the
temperature range of the measurements used in this work. The
advantage of our preparation method lies in the easy way to do
(one rubbing process is enough to produce samples, and from the
initial flake is possible to produce ten of substrates containing
multigraphene samples), it avoids contamination or surface doping
avoiding the contact with materials as with the Scotch-tape
method. For the transport measurements the substrate with the
sample was glued on a chip carrier using GE~7031 varnish. The
contacts from the chip carrier to the electrodes on the sample
substrate were done using a $25~\mu$m gold wire fixed with silver
paste.

\subsection{Transmission electron microscopy and electron backscattering diffraction}

The scanning electron microscope (SEM) pictures, electron
beam lithography, lamella preparation and electron backscattering
diffraction (EBSD) of the investigated samples
were done using a Nova NanoLab 200 dual beam
microscope from the FEI company (Eindhoven). A HOPG
lamella was prepared for transmission electron microscopy
(TEM) using the in-situ lift out method of the microscope. The TEM
lamella of HOPG was cut perpendicular to the graphene layers.
The electron transmission parallel to the graphene layers provides
information on the crystalline regions and their defective parts
parallel to the graphene layers. We
obtained thin lamellas of around 200~nm thickness, $\sim 15~\mu$m length
and $\sim 5~\mu$m width. After final thinning, the sample
was fixed on a TEM grid. A solid-state scanning transmission
electron microscopy (STEM) detector for high-resolution analysis
(included in our microscope) was used. The
voltage applied to the electron column was up to 30~kV and the currents
used were between 38~pA to 140~pA.

\begin{figure}[t] 
\centering
\includegraphics[width=8.5cm]{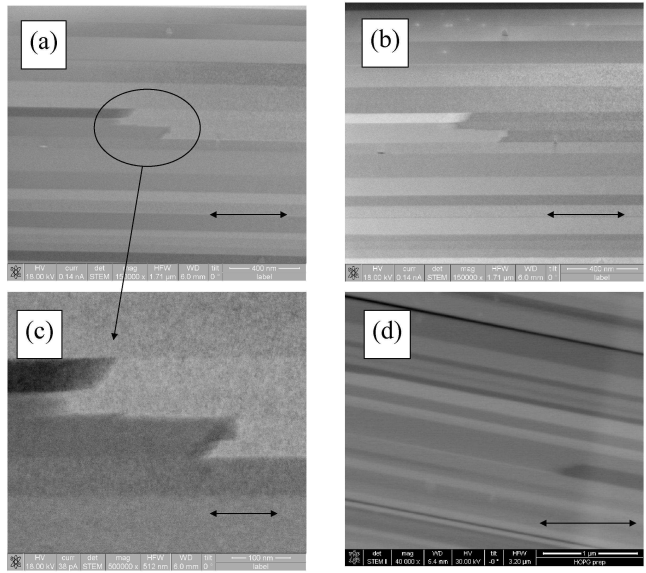}
\caption{Transmission Electron Microscopy pictures taken parallel to
the graphene layers of two HOPG lamella. The c-axis is perpendicular
to the clearly observable stripes of different gray colours, each
representing a crystalline region with a slightly different
orientation. The arrows in (a) and (b) indicate 400~nm length scale,
 in (c) 100~nm and in (d) 1~$\mu$m. Adapted from \protect\cite{bar08}.}\label{TEM1}
\end{figure}

Figure~\ref{TEM1} shows the bright (a) and dark field (b) details
obtained with the low-voltage STEM at 18~kV and (d) the bright
field picture of a different lamella obtained at 30~kV.
Figure~\ref{TEM1}(c) shows a blow out of a detail of (a). The
different gray colours indicate crystalline regions with slightly
different orientations. The images indicate that the average
thickness of the crystalline regions is $60 \pm 30~$nm. One can
clearly resolve the interfaces running perpendicular to the c-axis
of the layers, parallel to the graphene layers. In
Fig.~\ref{TEM1}(c) we also  realize the end of crystalline regions
along, normal or with a certain angle respect to the graphene
layers direction.

\begin{figure}[htb] 
\centering
\includegraphics[width=7cm]{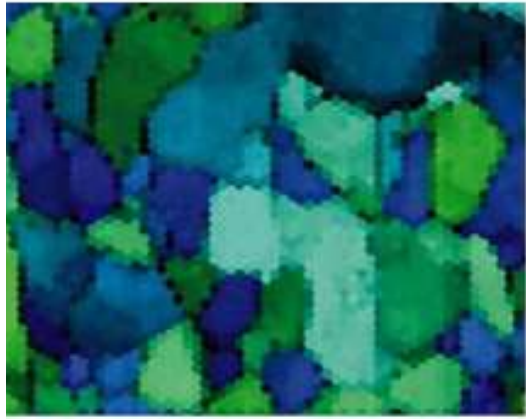}
\caption{ Electron backscattering diffraction pictures taken from
the surface of a  HOPG sample. The area scanned is $\simeq 80
\times 65~\mu$m$^2$ and the depth less than 10~nm.} \label{EBSD}
\end{figure}

Other experimental technique used to characterize the crystal
orientation, defects and grain boundaries is the electron
backscattering diffraction (EBSD). The measurements were performed
with a commercially available device (Ametek-TSL) included in our
microscope. In this setup a cleaved bulk HOPG sample under
investigation was illuminated by the SEM beam at an angle of
$70^\circ$ and the diffracted electrons were detected by a
fluorescence screen and a digital camera. The included software
was used to calculate the orientation of the crystalline regions
within the HOPG surface as a function of the electron beam
position. Figure~\ref{EBSD} shows the grain distribution at the
near surface region of a HOPG sample where the in-plane
orientation is recognized by the (bluegreen) colour distribution.
We recognize in this figure that the typical crystal size (on the
(a,b) plane) in our HOPG samples is between a few $~\mu$m and
$\sim 20~\mu$m. Taking into account the TEM pictures shown in
Fig.~\ref{TEM1} and the EBSD one in Fig.~\ref{EBSD} we conclude
that single crystalline regions in a HOPG good quality sample is
not more than $\sim 20~\mu$m long and less than $\sim 60~$nm
thick. If the interfaces between the crystalline regions as well
as the defects in the crystalline structure in graphite have some
influence on the transport properties we then expect to see a
change in the behavior of the transport properties between samples
of thickness of the order or less than the average thickness of
the crystalline regions and of planar size below
 $\sim 10~\mu$m.

\subsection{Influence of the internal microstructure on the temperature and magnetic field dependence of the longitudinal resistivity}\label{intmicro}

\begin{figure}[htb] 
\centering
\includegraphics[width=13cm]{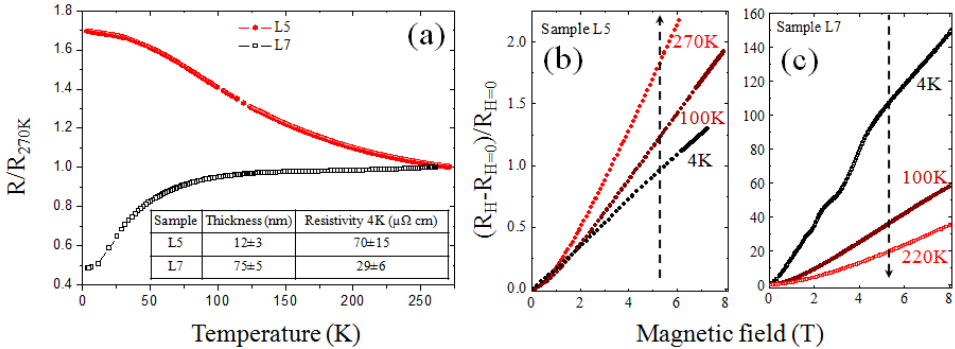}
\caption{(a) Normalized resistance versus temperature at zero
applied fields for two multigraphene samples, namely: L5 (red
dots) and L7 (empty black squares). The values of thickness and
resistivity at 4~K are included in the inset table. The
magnetoresistance (defined as shown in the y-axis) at three
different temperatures are shown for samples L5 (b) and L7 (c).
The magnetic field was applied parallel to the c-axis. Adapted
from \protect\cite{bar08}.}\label{L5-L7}
\end{figure}

In order to investigate the intrinsic properties of graphite and
taking into account the internal structure of bulk graphite
samples, it appears obvious to study tens of nanometers thick
graphite samples. \cite{bar08} correlated the internal
microstructure and sample size with the temperature and field
dependence of the electrical resistivity. Whereas HOPG or graphite
samples with thickness larger than $\sim 50~$nm show a
metallic-like behavior in the electrical resistivity vs.
temperature, in the case of tens nanometer thick graphite samples,
it steadily increases decreasing temperature, see
Fig.~\ref{L5-L7}(a).

The true temperature dependence of the resistivity in an ideal,
defect-free multigraphene sample should be semiconducting-like, as
it is expected for an ideal semimetal with zero or small gap.
\cite{dus11} noted that the temperature dependence of several
multigraphene samples shows a semiconducting-like behavior with a
saturation or a weak maximum at low temperatures, this last
behavior due to the contribution of interfaces, internal as well
as with the substrate or at the free sample surface, parallel  to
the graphene layers, see also Section~\ref{super}. It becomes now
clear that the contribution of the interfaces to the measured
conductance of large samples can overwhelm the one coming from the
intrinsic graphene layers in thick enough samples. Recently done
studies on several graphite samples of different thickness
indicate the existence of an intrinsic energy gap of the order of
40~meV in the graphene layers within ideal Bernal graphite
\citep{gar11}. We note that experiments on samples with carrier
concentration $ \lesssim 10^9~$cm$^{-2}$ are hardly reported in
literature. At such low-carrier densities, as it appears to
manifest in graphite, electron correlations and possible
localization effects should be considered. Electron interactions
are large and for a small-enough carrier density, the expected
screening will be very weak promoting therefore the existence of
an energy gap \citep{dru09}.

According to literature the structural quality of graphite samples
can be partially resolved by investigating the Raman D-line at
$\sim 1350~$cm$^{-1}$. However, also the edges of the samples as
well as the borderlines between regions of different thicknesses
may contribute to the D-band signal. The Raman spectra between
1300 and 1700 cm$^{-1}$ have been measured at different positions
of several multigraphene samples by \cite{bar08}. Those results
show that in thin multigraphene samples with similar
semiconducting-like behavior in the temperature dependence of the
resistivity,  different amplitudes of the Raman D-line peak are
measured. This indicates that the disorder related to this Raman
line does not appear to affect strongly the temperature dependence
of the resistivity.

As the absolute value of the electrical resistivity depends also
on the measured sample thickness, see Section~\ref{resthik}, it
seems clear to assume that the metallic-like behavior is not
intrinsic to ideal graphite but it is due to the influence of the
interfaces inside the graphite samples with large enough
thickness. Magnetoresistance (MR) results follow the behavior
observed in the electrical resistivity. If a metallic behavior is
obtained (thicker samples) then the MR decreases with increasing
temperature and it shows SdH oscillations, see
Fig.~\ref{L5-L7}(c). If the graphite sample shows a
semiconducting-like behavior (thinner samples) then the MR is
clearly smaller in magnitude as well as it {\em decreases}
decreasing temperature for fields above 1~T, see
Fig.~\ref{L5-L7}(b). This figure also shows the absence of the SdH
oscillations when the thickness of the graphite sample is small
enough.

We note that Kohler's rule does not apply in multigraphene and
HOPG samples and one of the reasons might be the contribution of
the sample internal structure and interfaces. Defects and
interfaces may influence the dimensionality of the transport and
might be the origin of localized granular superconductivity.
Results presented so far correspond to the situation in which the
field is parallel to the c-axis. In the case of having magnetic
field applied parallel to the graphene planes nearly no MR is
observed. This fact speaks for a huge anisotropy of the metallic
or superconducting regions and it suggests that those are within
the interfaces found by TEM, see Fig.~\ref{TEM1}. We note that the
field-induced metal-insulator transition (MIT) found in bulk HOPG
samples, see Section~\ref{bulkgra}, vanishes for samples thinner
than 50~nm, which also indicates that certain regions parallel to
graphene planes are related to the origin of this MIT
\citep{bar08}.

\subsection{Thickness dependence of the
resistivity}\label{resthik}

\begin{figure}[t] 
\centering
\includegraphics[width=8cm]{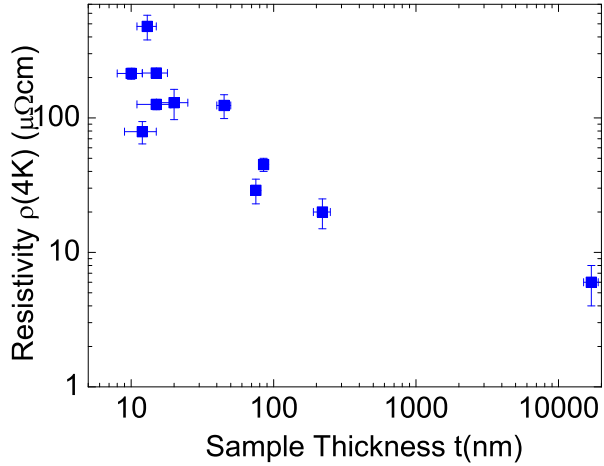}
\caption{Resistivity at 4~K vs. sample thickness of different
samples obtained from the same HOPG bulk sample.}\label{res-thick}
\end{figure}

Figure~\ref{res-thick} shows the resistivity at 4~K of different
samples vs. their thickness. It is clearly  seen that the
resistivity decreases increasing the sample thickness. The average
change in resistivity between $\sim 10~$nm
 to $17~\mu$m thick samples is  about two orders of magnitude, far beyond
 geometrical errors.  A
possible explanation for the observed behavior
 is that the larger the thickness the larger the amount of interfaces in the sample,
 see Fig.~\ref{TEM1}. As we described in Section~\ref{intmicro},
 there is a clear change in the temperature and magnetic field dependence decreasing
 the sample thickness. It appears unlikely, however, that randomly distributed
 point-like lattice defects can be the reason for the observed behavior.
Also the interpretation provided by \cite{kim05} that the decrease
of mobility $\mu$ (i.e. an increase in the resistivity at constant
carrier density) decreasing sample thickness provides an evidence
for boundary scattering is surely not the correct explanation for
the observed behavior, taking into account the fact that one
graphene layer shows finite and large mobility. Note that thin
graphite flakes show ballistic transport with huge mobilities, see
Section~\ref{ballistic}. The overall
 results suggest the existence of  a  kind of thickness-threshold  around $\sim 50~$nm
 for multigraphene samples of few micrometers size,
 obtained from HOPG bulk graphite indicating that
 the interfaces contribute substantially and in parallel to the graphene layers.

In the past the influence of the internal interfaces in graphite
bulk samples was either completely neglected or the scientific
community was not aware of their existence. On the other hand it
is well known that grain boundaries in various semiconductors can
lead to the formation of quasi-two dimensional carrier systems
confined in the boundaries. As early examples in literature we
refer to the quasi-two dimensional electron gas system that was
found at the inversion layers of n-Ge bicrystals
\citep{vul79,uch83} or in p-InSb \citep{her84} as well as in
Hg$_{1-x}$Cd$_x$Te grain boundaries \citep{lud92}. It is important
to note that in Ge-bicrystals \cite{uch83} found actually the
quantum Hall effect (QHE) at $T \leqslant 4.2~$K and at magnetic
fields above 6~T. The density of carriers at the interface was
estimated to be $\sim 5 \times 10^{12}~$cm$^{-2}$. Therefore, we
note that the usually reported carrier concentrations for graphite
are not intrinsic of ideal graphite, as shown by \cite{arn09} and
also by \cite{dus11}. Furthermore, we speculate that the QHE
behavior observed in bulk graphite samples
\citep{yakovprl03,kempa06} comes from internal interfaces.

\section{Searching for the intrinsic transport properties of the graphene layers inside graphite} \label{ballistic}

\subsection{Background}
In this Section we will briefly discuss the background of the
ballistic transport and its experimental observation in graphene
layers inside the graphite structure. The wave nature of the
electrons plays an important role when the sample dimensions are
comparable with the wavelength of the electrons. This turns to be
possible in graphite samples because the density of carriers is
very small, increasing therefore the Fermi wavelength $\lambda_F$.
On the other hand, as we will see below, a carrier within the
graphene layers inside graphite can transit micrometers through
the sample without collisions. Therefore the carriers in the
graphene layers within an ideal graphite structure have the
unusual property of having large $\lambda_F$ as well as large mean
free path $\ell$.

The transport phenomenon in the ballistic regime is best described
by the Landauer approach. Consider a narrow constriction connected
through two wide contact leads across which a voltage is applied.
Let the electrochemical potentials of two regions be $\mu_1$ and
$\mu_2$. The net current flowing through the device can be taken
as $I= (2e^2/h) \sum T_i (\mu_1-\mu_2)$ where $T_i$ is the
transmission probability of carriers and the factor 2 comes from
the spin degeneracy in each sub-band \citep{lan57,tsu73,but86}.
The effective conductance through a narrow constriction is given
by $G= I/(\mu_1-\mu_2) = (2e^2/h) \sum T_i$. When the constriction
dimensions are comparable to $\lambda_F$ and much shorter than the
mean free path $\ell$, one can see steps in the conductance when
the constriction width is varied. These steps are nearly integer
multiples of $(2e^2)/h$. Conductance quantization can be also
achieved by varying the potential energy. As the constriction
width is reduced or the gate voltage is changed in a determined
direction, the number of propagating modes at a given energy
decreases, i.e. the sub-bands are cut off one by one, and
therefore the conductance decreases. A new mode appears in the
conductance when the constriction width increases by
$\lambda_F/2$.

Nowadays, it is feasible to fabricate devices that show ballistic
transport. However, it is difficult and time consuming to
fabricate these devices because of the required small dimensions
when common metals or semiconductors are used. In the past there
were some experiments reveling conductance quantization at low
temperatures. The first experimental observations of conductance
quantization in two dimensional electron gases (2DEG) were
reported by \cite{wee88} and \cite{wha88}. They studied the point
contact conductance in GaAs-AlGaAs heterostructures as a function
of a negative gate voltage at low temperatures. The conductance
showed clear plateaus at integer multiples of $2e^2/h$ as the
width increases by an amount of $\lambda_F/2$. In both cases the
authors could show that the transport is completely ballistic and
the conductance is determined by the number of occupied sub-bands,
independently of the channel length.

\subsection{Ballistic transport in graphite}
For decades, the transport properties of bulk graphite were
interpreted using Boltzmann-Drude approach assuming diffusive
ohmic behavior for the conduction electrons \citep{kelly}. Within
this approach one has four temperature-dependent free parameters,
namely two mobilities and two carrier densities. The determination
of these parameters using this model implies fitting transport
data with at least four free parameters. To obtain accurate values
for these parameters it is necessary, however, to go beyond this
model. When the mean free path and Fermi wavelength are of the
order of sample size, one is not allowed to use the
Boltzmann-Drude transport theory to determine the electrical
resistance. As discussed before if the size of the system is of
the order or smaller than the carrier mean free path, ballistic
regime enters in which carriers can move through the system
without experiencing any scattering. Usually in metals this takes
place in the nanometer range. However, graphite is extraordinary
because its mean free path is of the order of microns
\citep{gar08,dus11}.

\begin{figure}[htb] 
\centering
\includegraphics[width=10cm]{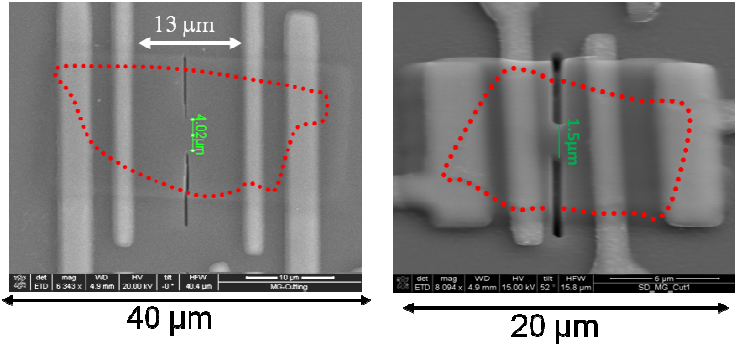}
\caption{Scanning electron microscope pictures of the two
multigraphene samples,  A (left) and B (right) with the Pd
electrodes. Sample~A has a constriction of $4\mu$m and sample~B of
$1.5~\mu$m. The scale bar in sample A indicates $10~\mu$m and in
sample~B $5~\mu$m. The dashed red lines denotes the border of the
multigraphene samples.}\label{cong}
\end{figure}

There is a transparent method to obtain all the main transport
parameters from experiments based on constrictions prepared in the
graphite or multigraphene samples \citep{gar08}. When the mean
free path is larger than the constriction width $(W)$, the
transport of the carriers shifts from ohmic to ballistic
\citep{sha65}. By measuring the longitudinal resistance as a
function of the constriction width, one can determine the mean
free path, carrier density and Fermi wavelength
\citep{gar08,dus11}.

Consider the geometry depicted in Fig.~1 of \cite{gar08}, or the
samples in Fig.~\ref{cong}, for a constriction in a quasi-two
dimensional sample. Already Maxwell pointed out that a
constricting circular orifice of diameter $W$ in a thin,
nontransparent screen of size $L_s$ produces a spreading
resistance that is equal to the $\rho(T)/W$ if the ratio $\ell/W
<< 1$ ($\rho$ is the resistivity of the material). However, when
this ratio increases there are two corrections to the Maxwell
spreading resistance: (i) the Ohmic value is corrected by a factor
of the order of unity as pointed by \cite{wex66}, and (ii) more
important, a dominant ballistic term appears. This was observed by
\cite{knu34,sha65,gar07} and the value of the resistance $R$ in
three dimensions reads \citep{wex66,gar07}:
\begin{equation}
 R_{3D}= \frac{4\rho(T)\ell}{3A} +
 \frac{\gamma(\kappa)\rho(T)}{W}\,,
 \label{wex-for}
 \end{equation}
where $A=\pi W^2/4$ is the area of the hole or constriction and
$\gamma(\kappa)$ is a smooth varying geometrical function, i.e.
$\gamma(\kappa = W/\ell)\simeq 1-0.33/\cosh(0.1\kappa) =0.67\ldots
1$ for $k=0\ldots \infty$ \citep{wex66}. In Eq.~(\ref{wex-for})
the first and the second terms of the right-hand side (rhs)
correspond with the Knudsen-Sharvin and Ohmic resistances,
respectively. The spreading Ohmic resistance in three dimensions
can be estimated within a factor $2/\pi$ off from the exact
Maxwell's solution assuming a hemisphere in which the electric
field $E(r)= J_{3D}(r)\rho(T)$. The radius $r$ is taken at the
constriction middle point and $J_{3D}(r)$ is the current density
equal to the total current $I$ divided by half of a sphere, i.e.
$J_{3D}(r)=I/(2\pi r^2)$, assuming that due to symmetry the
current is radial far away from the constriction.

From a similar calculation but in the two dimensional case,
appropriate for graphite due to the weak coupling between the
graphene planes, we have \citep{gar08}:

\begin{equation} R_{2D}(T)= a \frac{\rho(T)}{4Wt}
\ell(T) + a \frac{2\rho(T)\gamma(\kappa) ln(\Omega/W)}{\pi t}|_{W
< \Omega} + \frac{\rho(T)L}{Wt}\,. \label{r2d}
\end{equation}

The first term at the right-hand side (rhs) of Eq.~(\ref{r2d})
corresponds to the ballistic Knudsen-Sharvin resistance; the
second, logarithmic term to the ohmic, spreading resistance in
two-dimensions; here $\Omega$ is the total sample width and $t$
its thickness. The logarithmic dependence on the constriction
width of this ohmic, diffusive contribution is due to the
quasi-two dimensionality of the transport in graphite and supports
the assumption of weakly interacting graphene layers inside the
sample. The last term is due to the ohmic resistance of the
constriction tube itself. From previous works it was clear that
the position and shape of the voltage electrodes affects the
outcome of the experiment in mesoscopic devices \citep{mcl91}.
Therefore the constant $a$ was introduced, which takes care of the
influence of the sample shape, the topology, and the location of
the electrodes in the sample. For the usual electrode positions
through the whole sample width as shown in Fig.~\ref{cong}, $a =
1$. The validity of Eq.~(\ref{r2d}), especially the logarithmic
dependence of the ohmic part,  for HOPG as well as for
multigraphene samples has been verified by \cite{gar08} and
\cite{dus11}.

In the following, we review some of our experimental results for
two multigraphene samples, A and B, with different geometry and
resistivity. The sample details as well as their preparation and
fabrication were described by \cite{bar08,dus11}. The
constrictions in the middle of the samples, see Fig.~\ref{cong},
were prepared with the focused-ion beam of a dual-beam microscope.
It is important to note that we avoided the modification of the
crystalline structure of the samples due to the ion beam spread by
protecting them with a $\sim 300~$nm thick negative-electron beam
resist (AR-N 7500), a method successfully tested by
\cite{barnano10}.

\begin{figure}[htb]   
\centering

\subfigure[Sample A at $T = 60~$K]{\includegraphics[width=6cm]{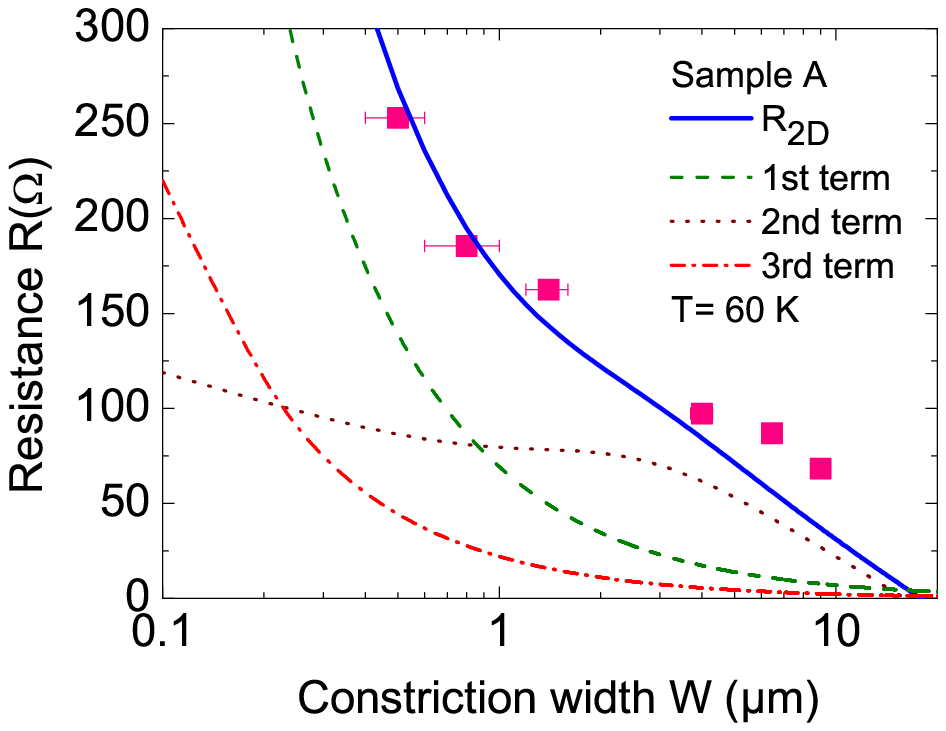}} 
\subfigure[Sample A at $T = 250~$K]{\includegraphics[width=6cm]{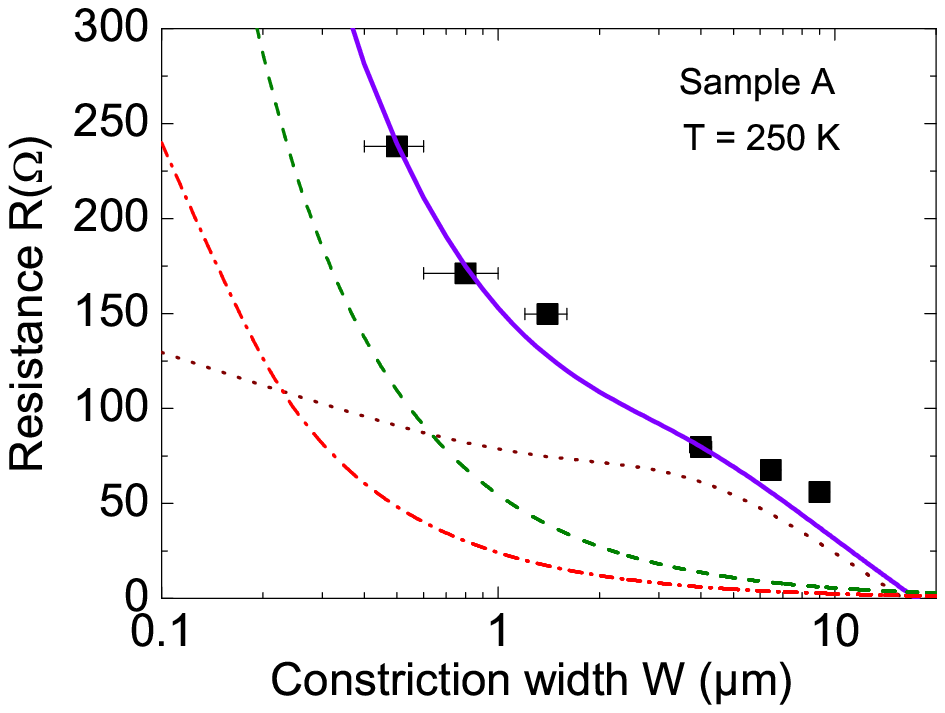}} 
\subfigure[Sample B at $T =
10~$K]{\includegraphics[width=6cm]{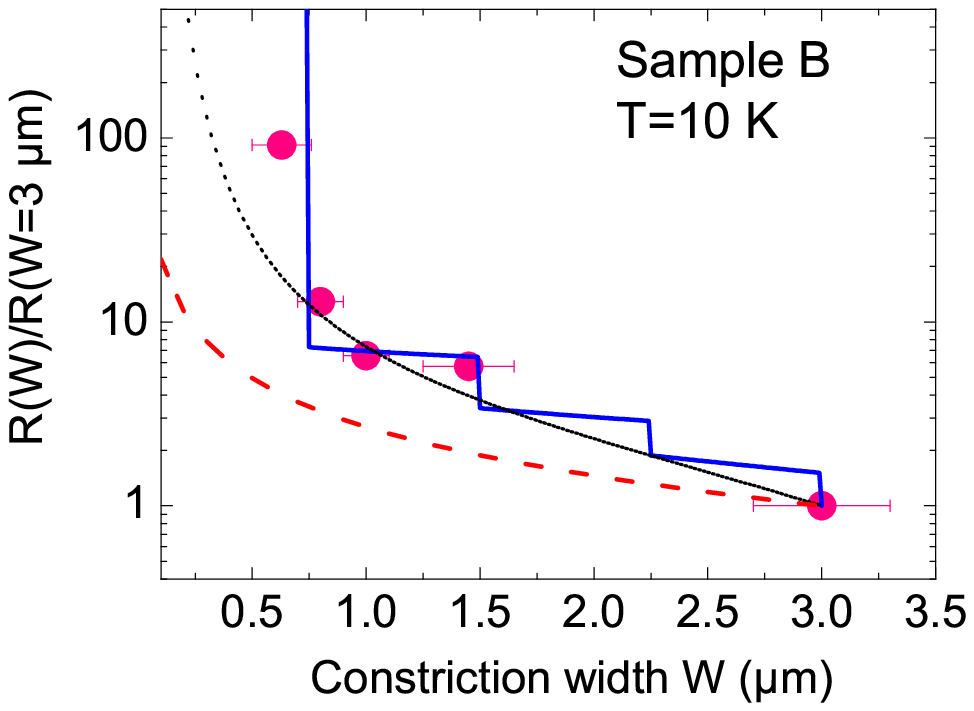} } \subfigure[Sample B
at $T = 300~$K]{\includegraphics[width=6cm]{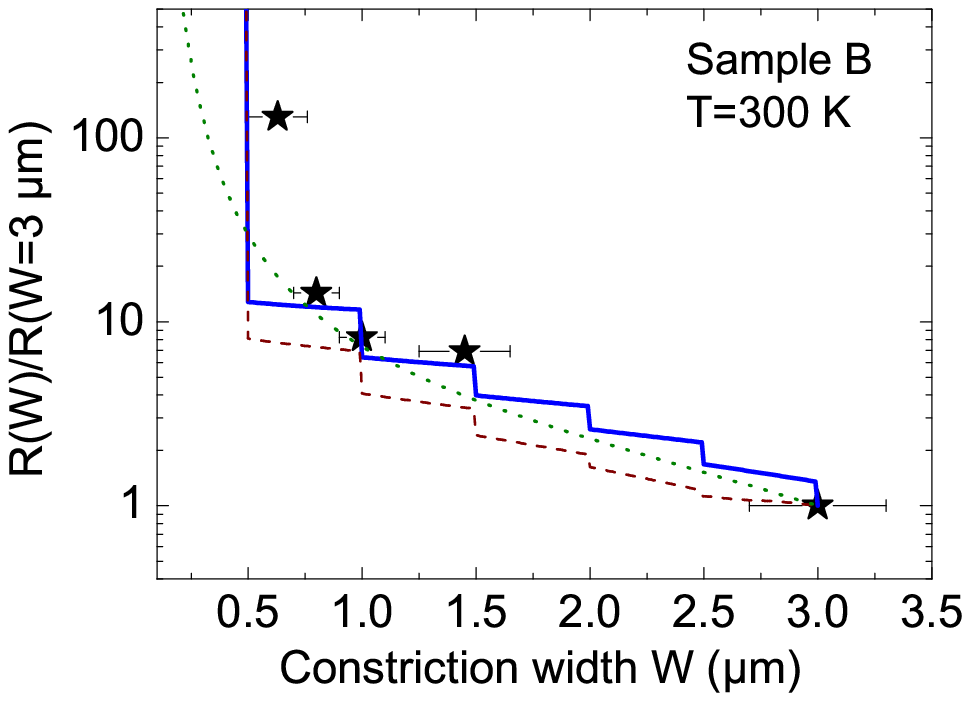} }

\caption{Measured resistance as a function of the constriction
width $W$ at (a) 60 K and (b) 250 K for sample A. The point with
the largest $W$ corresponds to the virgin sample without a
constriction. The different lines correspond to the ballistic
contribution (dashed line) and  ohmic second (dotted line) and
third (dashed-dotted line) terms in Eq.~(\protect\ref{r2d}). The
continuous line represents the addition of the three
contributions. For (a) the continuous line is calculated with
$\ell = 1.2~\mu$m and for (b) $\ell = 0.8~\mu$m. (c) and (d):
Normalized resistance for sample~B vs. constriction width $W$ at
10~ and 300~K. Note the semi logarithmic scale. The  line  with
steps  is  obtained  dividing  the  ballistic  term in
Eq.~(\protect\ref{r2d}) by $(\lambda_F /2W){\rm
trunc}(2W/\lambda_F)$ with the parameters $\ell = 2.7~\mu$m and
$\lambda_F  = 1.5~\mu$m. The dashed line follows
Eq.~(\protect\ref{r2d}). The dotted line is obtained multiplying
the ballistic term in Eq.~(\protect\ref{r2d})
 by the exponential function $\exp(\lambda_F/2W)$.
(d) The same as in (c) but the continuous line was obtained with
the parameters $\lambda_F = 1.0~\mu$m and $\ell = 2.2~\mu$m. The
dashed stepped function is obtained using the same $\lambda_F$ but
with a smaller $\ell = 1.3~\mu$m. Adapted from \cite{dus11}.}
\label{rvsw}
\end{figure}

To obtain the mean free path without further adjustable parameter
we measured the resistance $R$ as a function of the constriction
width $W$ and use Eq.~(\ref{r2d}). Figure~\ref{rvsw} shows the
results for samples~A and B at two temperatures.  The results show
that for $W < 1~\mu$m the ballistic contribution (dashed lines in
Figs.~\ref{rvsw}(a) and (b)) overwhelms the ohmic ones indicating
that the mean-free path should be of this order. Having only
$\ell$ as free parameter Eq.~(\ref{r2d}) can be used to fit the
behavior of $R$ vs. $W$ for sample~A. From the theoretical fits
one obtains $\ell = 1.2~\mu$m and $0.8~\mu$m at 60~K and 250~K,
respectively. The Fermi wavelength per graphene layer can be
calculated using \citep{gar08}:
\begin{equation}
\lambda_F = \frac{1}{a_0} \frac{2\pi \rho(T) \ell(T) e^2}{h}\,,
\label{lf}
\end{equation}
where $a_0 = 0.335~$nm is the distance between graphene planes in
the graphite Bernal stacking configuration. For sample A we obtain
then $\lambda_F = 0.5(0.8) \pm 0.25~\mu$m at 250~K (60~K).

Despite of the good agreement obtained for sample~A and for HOPG
bulk samples \citep{gar08}, Eq.~(\ref{r2d}) suffers from an
important limitation since it cannot describe correctly ballistic
transport phenomenon in which the wave nature of the electrons
plays a crucial role, i.e. in samples where $\lambda_F \gtrsim W$.
As discussed before, in this case the ballistic contribution to
the resistance is better described by the quantization of the
transverse electron momentum in the constricted region. In this
case the value of the resistance is given by the inverse of a sum
of an energy-dependent and transverse wave vectors $q_n$-dependent
transmission probabilities $T_n$, where $n = 0, \pm 1, \pm 2,
\ldots N_c$ \citep{sta07}. These values correspond to the one
dimensional electric sub-bands and $N_c$ is the largest integer
smaller than $2W/\lambda_F$. In this case the increase in
resistance is expected to show an oscillatory behavior as a
function of $W$ or $\lambda_F$ \citep{gar89,sny08} as observed
experimentally in bismuth nanowires \citep{cos97} as well as in
GaAs devices \citep{wee88,wha88}.

Note that the obtained mean free path for sample~A is smaller than
the distance between the electrodes. The larger the sample, larger
is the probability to have defective regions with larger carrier
concentration and smaller mean free path within the voltage
electrodes \citep{arn09}. In order to increase the probability to
observe the expected quantization phenomenon in multigraphene
samples, it is necessary to have a mean free path larger than the
sample size in order to be completely in the ballistic regime.
Therefore, we repeated the experiment with sample~B that shows
lower resistivity and with a smaller voltage-electrode distance,
see Fig.~\ref{cong}.

Figures \ref{rvsw}(c) and (d) show the measured resistance
normalized by its value at a constriction $W = 3~\mu$m for sample
B. The normalization is necessary because in this way we pay
attention to the huge relative increase of $R$ decreasing $W$ and
we need neither the absolute value of $\rho$ nor of $a$ to compare
the data with theory. We realize that for sample B Eq.~(\ref{r2d})
does not describe accurately the experimental data even assuming
the largest possible mean-free path equal to the voltage-electrode
distance of $\simeq 2.7~\mu$m. The data can be reasonably well
fitted dividing the ballistic term in Eq.~(\ref{r2d}) by the
function ${\rm trunc}(2W/\lambda_F)\lambda_F/2W$, which generates
steps whenever the constriction width $W \simeq i\lambda_F/2$ with
$i$ an integer. From the fits we obtain the parameters $\lambda_F
= 1.0(1.5)~\mu$m and $\ell= 2.2(2.7) \pm 0.3~\mu$m at 300(10)~K.
Using other values of $\ell$, for example $\ell = 1.3~\mu$m, see
Fig.~\ref{rvsw}(d), the function does not fit the data indicating
indeed that the carriers behave ballistically between the voltage
electrodes, leaving actually $\lambda_F$ the only free parameter.

The ballistic analytical function we used resembles the
theoretical results with similar steps obtained by \cite{sny08}
where the conductance vs. $W$ was calculated numerically for a
single layer graphene with an electrostatically potential
landscape that resembles a constriction. An analytical average
value or envelope of this stepped function is obtained replacing
the truncation function by $\exp(-\lambda_F/2W)$, see
Figs.~\ref{rvsw}(c) and (d). This exponential function represents
the impossibility of an electron to propagate in the constriction
when $W < \lambda_F/2$. The important result obtained for sample B
is the huge increase of the resistance for $W < 2~\mu$m indicating
clearly a larger $\ell$ than the one obtained in sample A, see
\cite{dus11} for further details.

\begin{figure}[htb] 
\centering
\includegraphics[width=6.5cm]{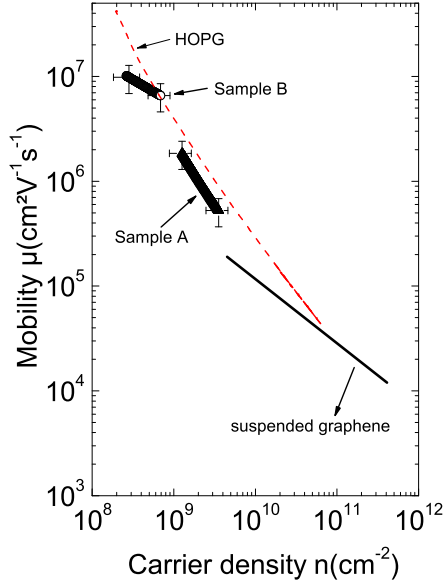}
\caption{Mobility vs. carrier density obtained for samples A and B
(after \protect\cite{dus11}) and for HOPG (dashed line, after
\protect\cite{gar08}). Note that the data points are not obtained
at a constant temperature but the temperature is changed to get a
change in the carrier density and mobility. The points are
obtained between 300~K and 10~K for sample B and 300~K and 60~K
for sample A. For the HOPG sample  (dashed line) the data run from
5~K at the largest mobility values to 300~K. The straight line
corresponds to the data of a suspended graphene sample at 20~K
from \protect\cite{du08}.}\label{muvsn}
\end{figure}

The  temperature dependence  of $R(T,W)$ can be used now to obtain
$\lambda_F(T)$ and the mobility per graphene layer, this last
given by $\mu(T) = (e/h)\lambda_F(T)\ell(T)$. Since  the  density
of carriers per graphene layer  can  be calculated from $n
=2\pi/\lambda_F^2$ we show in Fig.~\ref{muvsn} the mobility vs.
carrier density for the two samples and for a bulk HOPG sample and
compare them with data from literature for suspended single layer
graphene. From these results we clearly recognize the much larger
mobility and smaller density of carrier for the graphene layers
inside graphite, supporting the view that the graphene layers
within graphite are of better quality and with a smaller carrier
density than single layer graphene. Obviously neither sample A nor
B nor the HOPG sample are free from defects and therefore we
expect that the obtained values might still be improved in ideal,
defect-free graphite structures.

\section{The influence of interfaces inside bulk graphite samples: Hints for the existence of high-temperature superconductivity}\label{super}

\subsection{General remarks}
 Ideal graphite is a layered material where
each of the layers (graphene) consists on a honeycomb lattice of
carbon atoms. Graphene layers are stacked together with Van der
Waals forces much smaller than the covalent ones between the
carbon atoms within a single layer. This weak coupling between
graphene layers within the graphite structure can be "translated"
in a huge anisotropy in the resistivity $\rho_c / \rho_{a} \gtrsim
10^6$ at low temperatures. In high quality samples this leads to a
quasi-two dimensionality of the transport because most of it
occurs within the graphene layers inside the graphite structure
\citep{yakadv07}. However, real graphite cannot be considered as
an ideal uniform stack of ideal graphene layers, as we discussed
in Section~\ref{sample}. The role of defects
\citep{arn09,barzola2,barjsnm10} as well as internal interfaces
\citep{bar08} have a crucial effect in the electrical properties.
Clear evidence on the non uniformity of real graphite structures
is given by several microscopy techniques, see
Section~\ref{sample}. In this Section we are interested in
particular on the influence of the internal interfaces on the
transport.

\subsection{Bulk graphite}
\label{bulkgra}
 Starting with the bulk material, \cite{kempa00}
studied the magnetoresistance in the in-plane direction in HOPG
bulk  samples where a magnetic-field-driven transition from
metallic- to semiconducting-type behavior of the basal-plane
resistance $\rho_a$ was found. This was reproduced in later
publications by other groups \citep{heb05,tok04}. Note, however,
that depending on the thickness of the graphite samples and their
quality, metallic or semiconducting behavior in graphite can be
observed if no magnetic field is applied \citep{bar08}. There is a
clear suppression of the metallic-like phase (in case it appears)
by a magnetic field applied perpendicular to the graphene planes
(i.e. parallel to the c-axis) \citep{kempa00}. Later work found a
field-induced metal-insulator transition (MIT) also in the c-axis
resistivity of graphite $\rho_c$ \citep{kempa02}. The
metallic-like behavior of $\rho_c$ has been related to the one
found in the longitudinal resistivity due to a conduction-path
mixing mechanism, i.e. the conduction path of the carriers along
the c-axis is in part short circuited by lattice defects and/or
impurities.

If non-percolative superconducting grains are located at
interfaces running mostly parallel to the graphene planes, the
largest increase in the resistance with magnetic field applied
parallel to the c-axis is expected in the temperature region where
either the resistance decreases decreasing temperature or it
levels off at zero field applied. This is expected since at a
temperature below the maximum or leveling off one speculates that
the coupling between superconducting grains starts to be
observable. \cite{yakov99} showed a sensitive change in the
temperature dependence of the resistivity in bulk HOPG samples
under an applied magnetic field parallel to the c-axis. Additional
studies done by \cite{kempa03} show the absence of a MIT in
graphite if the magnetic field is applied parallel to the graphene
planes, a fact that indicates that its origin is related to
regions running parallel to the graphene layers.

\subsection{Anomalous hysteresis loops and quantum resonances in the magnetoresistance}\label{ahloops}

\begin{figure}[htb] 
\centering
\subfigure[Low field hysteresis]{\includegraphics[width=6cm]{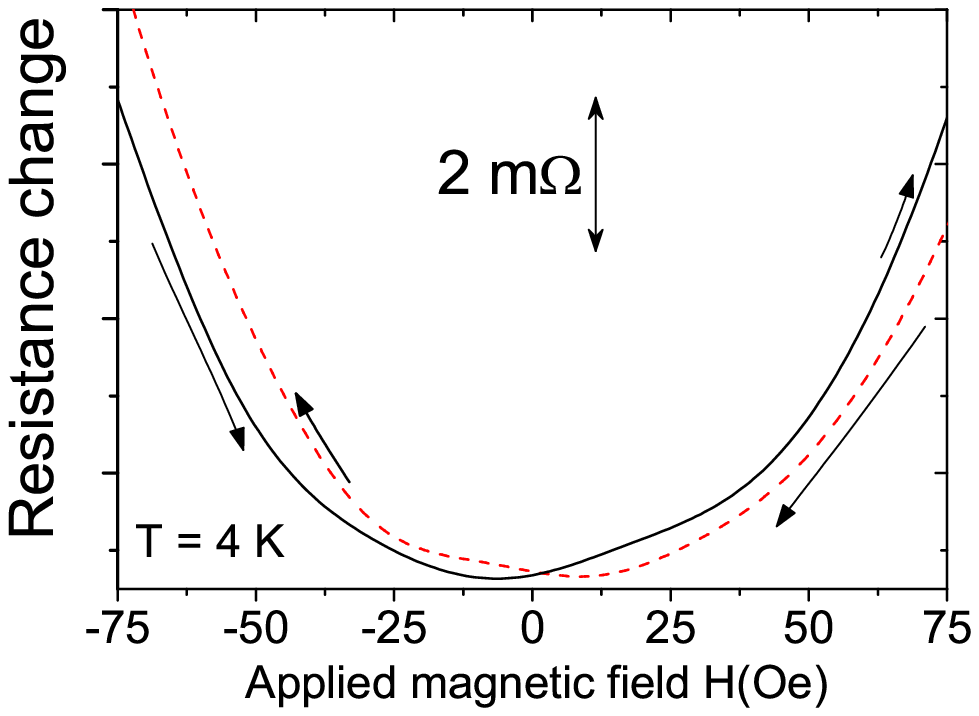}} 
\subfigure[Quantum oscillations]{\includegraphics[width=6cm]{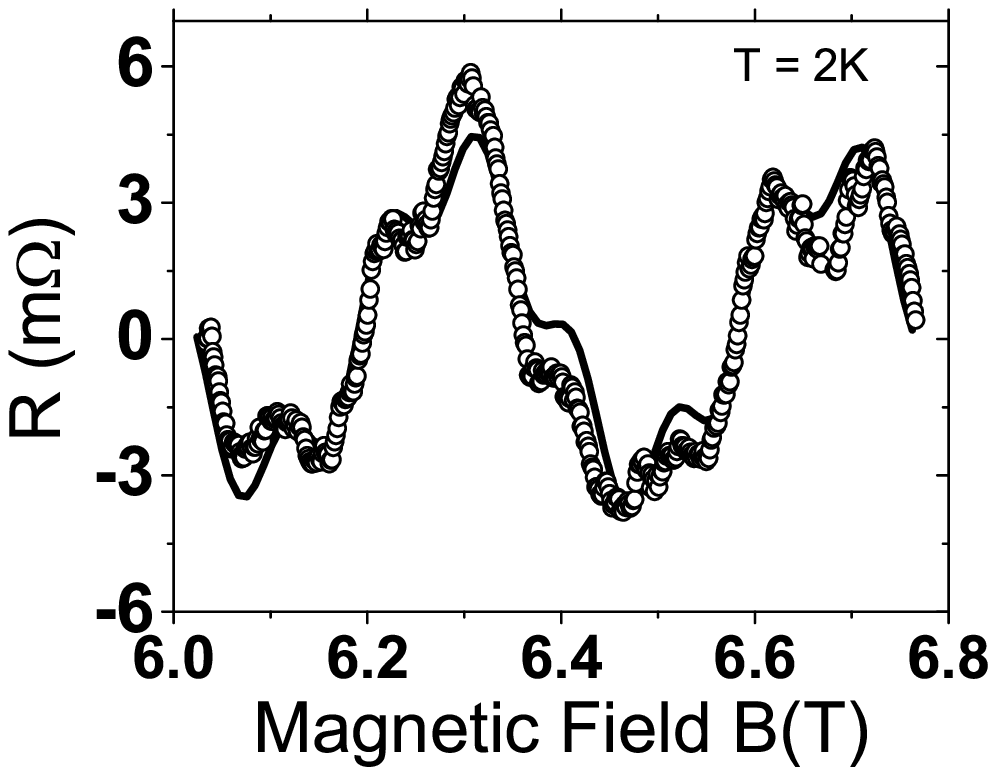}} 
\caption{(a) Weak hysteresis in the resistance of a thin graphite
sample vs. magnetic field, similar to that observed in
superconductors with Josephson-coupled grains near zero field,
measured in a multigraphene sample (after \protect\cite{esq08}).
 (b) Oscillations in the resistance
of a multigraphene sample as function of the magnetic field at 2~K
after subtraction of a linear field background. The continuous
line corresponds to a Fourier fit. Adapted from
\protect\cite{esq08}.}\label{AHL}
\end{figure}

In percolative homogenous superconducting samples, one can ascribe
superconductivity by observing a screening of the external
magnetic field (Meissner effect) below a critical field and/or by
measuring the drop of resistance to practically zero below a
critical temperature. Inhomogeneous granular superconducting
materials (in which the superconducting phase could be found in
small parts of the whole sample) require other way of testing, as
resistance does not  drop necessarily to zero and the Meissner
effect is probably too small to be measured. \cite{esq08} found
anomalous hysteresis loops in the MR similar to those observed in
granular superconductors with Josephson-coupled grains, see
Fig.~\ref{AHL}(a).

\cite{esq08} and \cite{sru11} show that the minima in the MR are
located in the same quadrant field from which one starts the field
sweeping as it was reported in conventional high $T_c$ granular
superconductors \citep{ji93}. This can be explained based on a
two-level critical-state model where pinned fluxons exist inside
the Josephson-coupled superconducting grains but also between
them. These last ones are usually much less pinned and therefore
can strongly influence the MR behavior, especially at low enough
fields. The fact that the minimum of the MR is rather flat can be
interpreted as due to non-uniform superconducting grain size.

In addition, it must be stressed that when the field is parallel
to the graphene planes no effect is observed, suggesting that the
superconducting regions or patches are rather parallel to the
graphene layers. Quantum resonances in the MR were also observed,
see Fig.~\ref{AHL}(b), and ascribed to Andreev's reflections
between localized superconducting patches connected by
semiconducting regions in which Cooper pairs may flow relatively
large distances \citep{esq08}. From the amplitude and period of
the oscillations in field, the distance between superconducting
granular domains was estimated to be $\lesssim 1~\mu$m
\citep{esq08}.

\subsection{Behavior of the magnetoresistance of multigraphene samples with  micro-constrictions}\label{mrcons}

\begin{figure}[htb] 
\centering
\includegraphics[width=8cm]{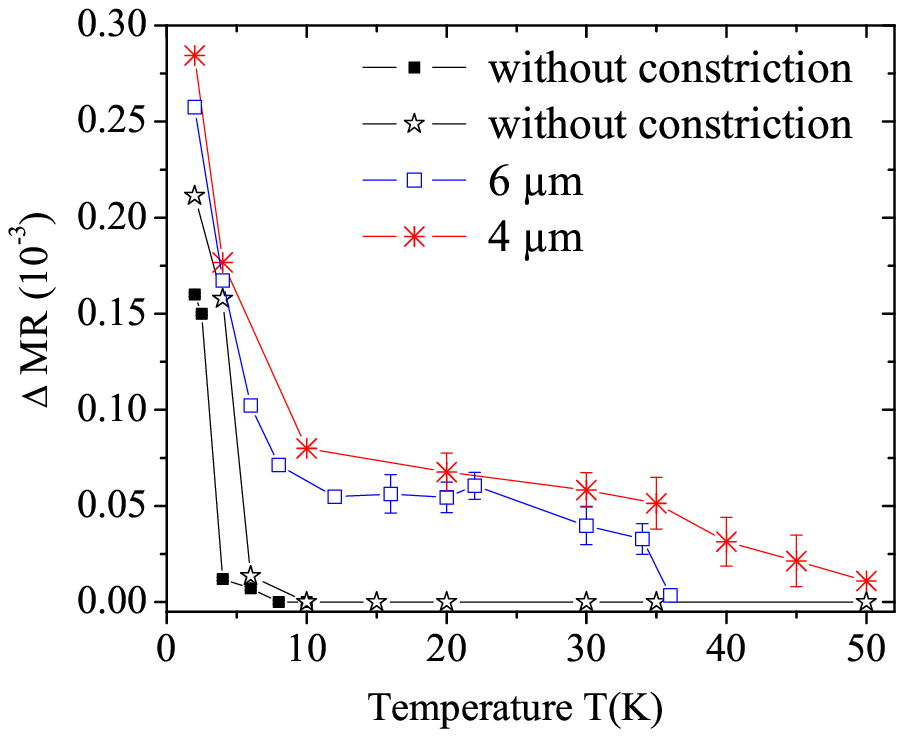}
\caption{Relative difference between the two field branches of the
hysteresis loop in the MR (see Fig.~\protect\ref{AHL}(a)) as a
function of temperature for two different multigraphene samples at
a constant field of 1.66~mT. Black squares $(\blacksquare)$
correspond to the data presented by \cite{esq08};
($\star,\Box,\ast$) correspond to the results presented by
\cite{sru11} for a multigraphene sample without and with two
constrictions. The increase in the hysteresis using the
constriction method is clearly seen. The samples without
constrictions show granular superconducting-like hysteresis loops
up to 10~K only, while samples with the constrictions show this
behavior up to 50~K in the case of $4~\mu$m constriction
width.}\label{DMR}
\end{figure}

The novel method of using thin mesoscopic samples with
micro-constrictions not only allows us to obtain basic parameters
like the mean free path and carrier density, see
Section~\ref{ballistic}, but provides also the possibility to
increase the sensitivity of the voltage measurement to the
superconducting regions that may be at the constriction region. We
mean that the expected superconducting patches inside the
constriction area might be better "detected" by usual voltage
measurement at two contacts on either side of the constriction.
\cite{sru11} verified this effect and found an increase in the
range of temperature where the previously reported \citep{esq08}
anomalous hysteresis loops can be observed, see Fig.~\ref{DMR}.

All the experimental results indicate that, in case that the
superconducting regions exist, they should be localized mainly at
the interfaces between the crystalline regions of slightly
different orientation observed by TEM, see Fig.~\ref{TEM1}. At
these interfaces the density of carriers should be high enough to
achieve high critical temperatures, provided the quasi-two
dimensionality remains \citep{garbcs09}. The sample can be modeled
as the sum of superconducting and normal in series and in parallel
circuit paths. A coherent superconducting state at low enough
temperatures within the superconducting patches is expected as
well as in between some of them. Thermal fluctuations influence
the coherent superconducting state in some parts of the sample and
no zero resistance is achieved. To summarize the results reported
so far in tens of nanometers samples as well as the first
indications in HOPG bulk samples, two fundamental aspects must be
remarked: first, metallic-like behavior of the resistivity in HOPG
bulk is not intrinsic and not related to the scattering of
conduction electrons with phonons, and second, if
superconductivity plays a role in the anomalous properties
observed in HOPG samples then the quasi-two dimensional interfaces
may contain the superconducting regions \citep{gar11}.

\subsection{Transport measurements of TEM lamellae}

\begin{figure}[htb] 
\centering
\includegraphics[width=12cm]{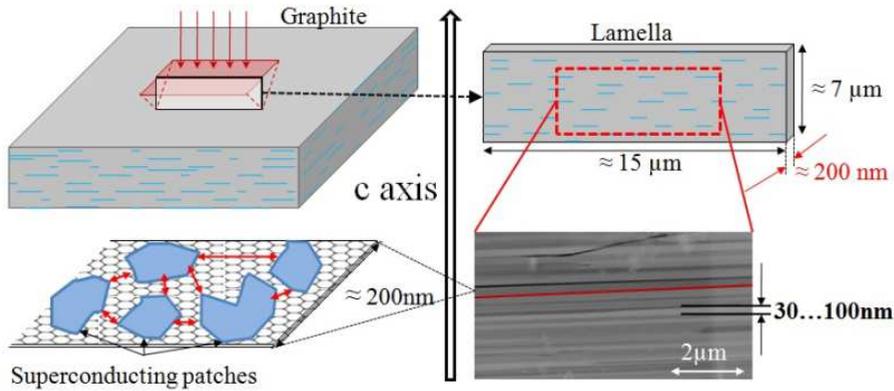}
\caption{Sketch of the lamellae preparation process. Focused Ion
Beam (FIB) is used to cut bulk HOPG (up left side: reddish areas
represent the area to be cut with Gallium ion beam and this last
is represented by the red arrows). Previous to this cut, the
surface of HOPG is protected with WO thick layer deposited using
the electron beam. Lamellae of size $\sim 7~\mu$m~$\times
15~\mu$m~$\times 200~$nm (this last dimension in the graphene
plane direction) are prepared as shown in the upper right draw.
The low-TEM picture (bottom right) shows the interfaces between
the crystalline regions (one of them marked in red). Blue lines
drawn in the upper pictures provide an idea where the quasi-two
dimensional superconducting patches are supposed to be located.
The bottom left picture represents how the situation at one of the
interfaces might be. The blue areas simulate the superconducting
patches that might be weakly, Josephson coupled by low-carrier
density graphene regions.}\label{lam}
\end{figure}

As explained before, the indications for superconductivity in
graphite get clearer when the sample dimensions are reduced to
some extent. On the other hand, if the sample dimensions are
further reduced, the intrinsic properties of the graphene layers
inside graphite can be investigated. In order to further
investigate the role of interfaces a different way of preparing
samples from bulk graphite seems necessary. The main idea is to
prepare small and narrow graphite samples (within the graphene
planes) in order to confine the path of the input current to go
through some of the possible superconducting interfaces running
mostly parallel to the graphene planes. Further details of the
process as well as the lamella dimensions can be seen in
Fig.~\ref{lam}. Several lamellae have been studied and the results
are presented in the following lines.

\begin{figure}[htb]   
\centering

\subfigure[Sample 1]{\includegraphics[width=6cm]{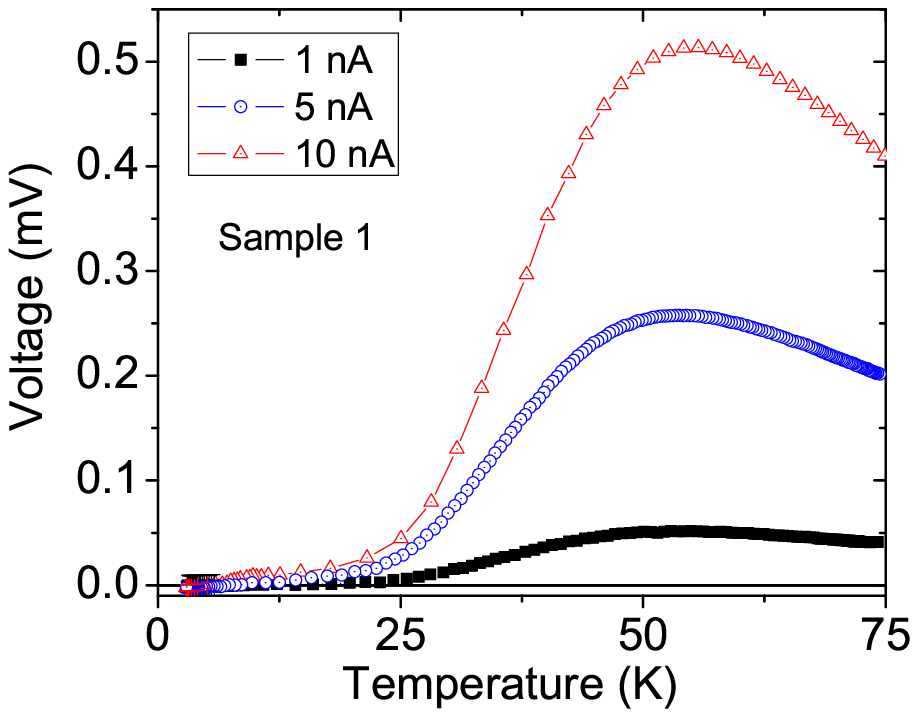}} 
\subfigure[Sample 2]{\includegraphics[width=6cm]{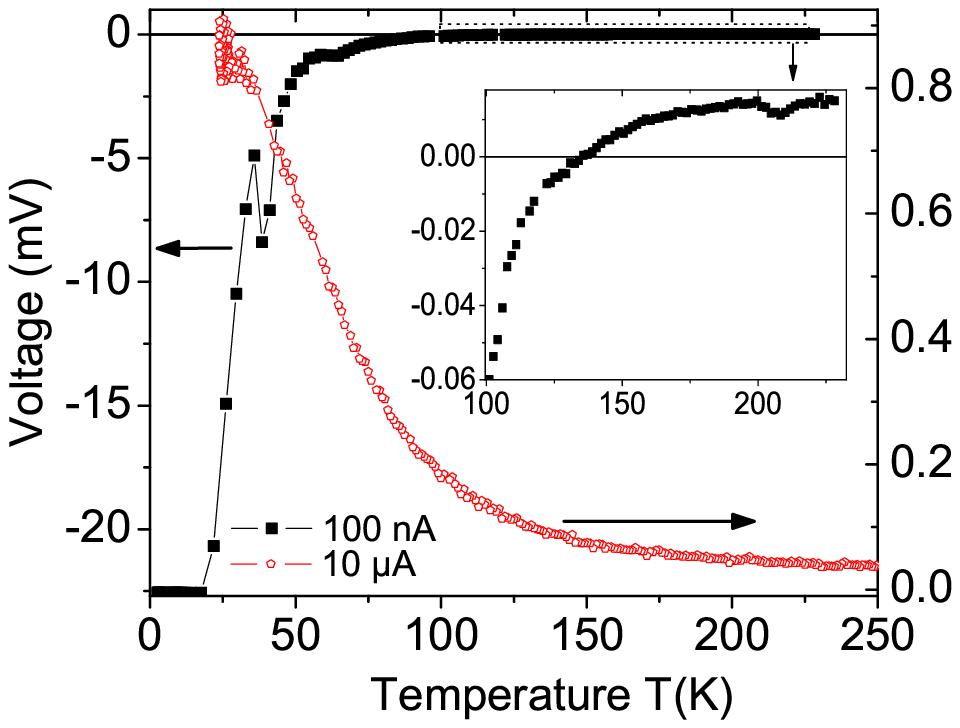}} 
\caption{Voltage vs. temperature measured for two different
lamellae. Note the clear change in the temperature dependence with
input current. A drop in the measured voltage is observed for all
lamellae when the input current is small enough.} \label{vvst}
\end{figure}

In Fig.~\ref{vvst} we show the change of the voltage vs.
temperature measured at two positions in two different lamellae,
samples 1 and 2, both prepared using the same procedure. The
estimated critical temperatures for these samples are: 25~K and
175~K, respectively, although we should clarify that these
temperatures not necessarily are the intrinsic critical
temperatures of the patches but probably those related to the
Josephson coupling between them. The differences observed in the
values of $T_c$ reflect the inhomogeneous structure expected for
the graphite interfaces.  Sample 1 results can be considered as
the "ideal ones" because within resolution zero values in the
voltage are reached for low enough temperatures and small input
currents, see Fig.~\ref{vvst}(a). In this sample, the used
electrical contacts for the voltage measurements were directly at
or near some superconducting paths that short circuited below
certain temperature and input current. Sample 2 shows similar
results, qualitatively speaking, but the range of critical values
in temperature and current are different. In this sample the
contacts are not placed directly on the superconducting regions
and therefore normal regions of the samples contribute to the
whole measured voltage. It can be seen in Fig.~\ref{vvst}(b) that
in this case the transition is sharper and a change of two orders
in magnitude in the measured voltage is achieved. Due to the van
der Pauw contact configuration used, negative values (respect to
the input current direction) of the voltage below the transition
temperatures are observed for several lamellae, instead of zero
voltage. This can be explained using a Wheatstone bridge circuit
as it will be shortly described later in this Section. We found a
strong dependence of the measured voltage on the input current.
Above a certain current neither negative nor the drop in voltage
are observed anymore.

\begin{figure}[htb]   
\centering

\subfigure[Sample 1]{\includegraphics[width=6cm]{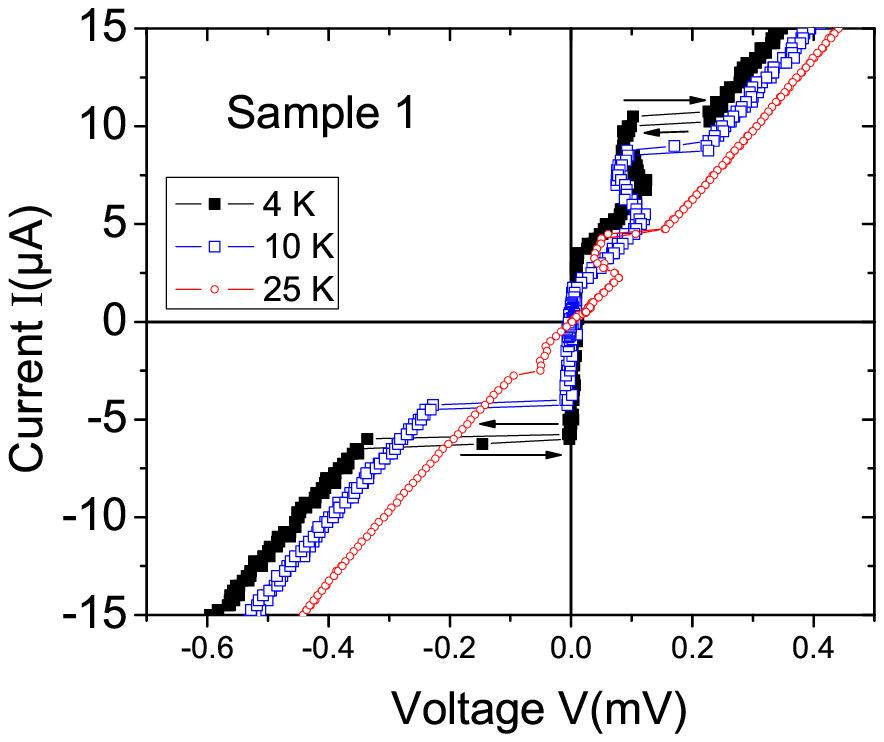}} 
\subfigure[Sample 2]{\includegraphics[width=6cm]{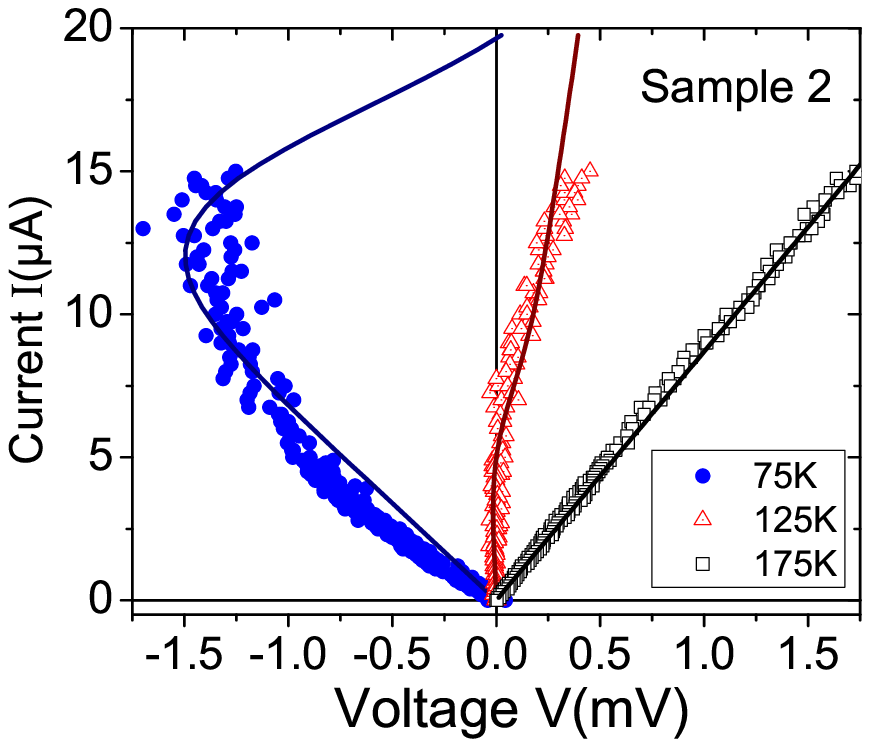}} 
\subfigure[Sample 3]{\includegraphics[width=6cm]{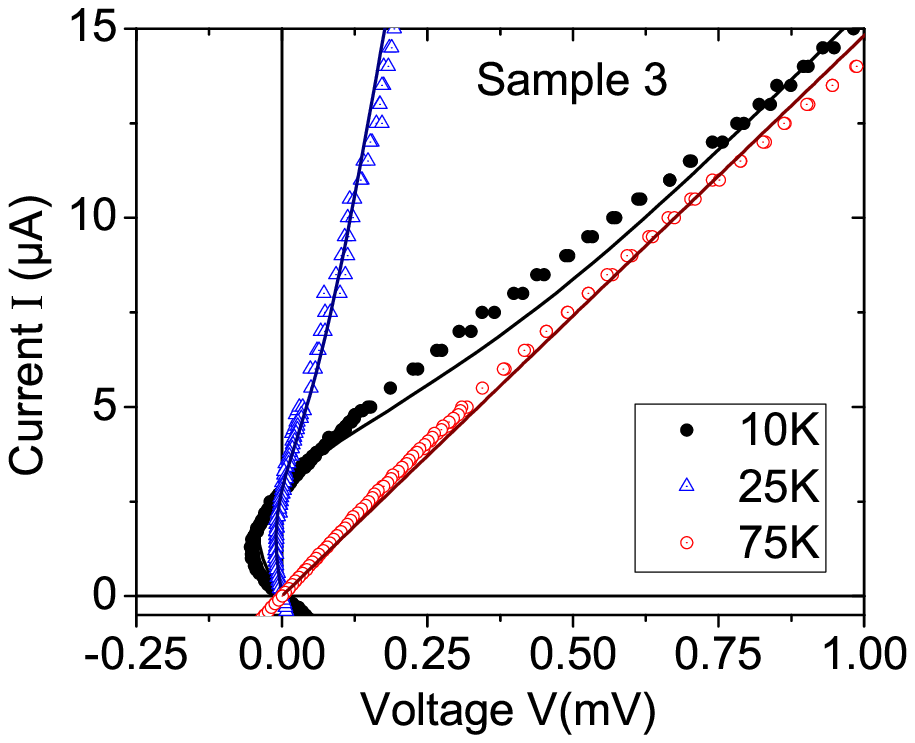} }

\caption{Current-voltage characteristic curves at different fixed
temperatures for three different lamellae. The continuous lines
that fit the data were obtained using the \protect\cite{amb69} CVC
for two $R_i$'s in (b) and one $R_i$ in (c) within a Wheatstone
bridge model, assuming different critical Josephson currents at
different temperatures. No magnetic field was applied and the
earth field was not shielded.} \label{cvc}
\end{figure}

Current-Voltage characteristic curves (CVC) are the next important
piece of evidence to be mentioned. In case of having Josephson
coupling the expected CVC should show a nearly zero voltage value
in the measured voltage below a certain value of current and at
low enough temperatures. At currents much above a critical one an
ohmic linear behavior should be recovered. Another important
feature that particularly at low temperature is found in
Josephson-coupled systems is the irreversibility in the CVC. As
shown in Fig.~\ref{cvc}(a)
 this is achieved in the case of the "ideal" sample 1. The CVC results for
other two lamellae become a bit more complex but compatible with
the previous results shown in Fig.~\ref{vvst}(b).
Figures~\ref{cvc}(b) and (c)  correspond to the CVC for samples 2
and 3, respectively. Negative values in the differential
resistance are measured below certain values of current and
temperature and above them a linear ohmic behavior is recovered.
The simplest way to explain these results is by using a
non-homogeneous current-voltage circuit composed by four different
resistors $R_i, i = 1 \ldots 4$ within a simple Wheatstone bridge
model as for example: $V(I) \propto I (R_1R_4 - R_2R_3)/((R_1 +
R_2)(R_3 + R_4))$. Each of the $R_i$ represents an effective
resistance within a path inside the lamella. In case one of the
resistors includes Josephson-coupled superconducting regions with
zero voltage value at low enough temperatures and currents, we can
simulate the measured behavior using for this $R_i$ the CVC given
by \cite{amb69}. \cite{amb69} described the voltage vs. current
behavior within the DC Josephson effect including thermal noise.

As it can be seen in Figs.~\ref{cvc}(b) and (c) the theoretical
calculations (solids lines) -- using one free parameter (the
critical Josephson current) to calculate the corresponding
$R_i(I)$ following \cite{amb69} and fixing two or three arbitrary
current-independent resistor values for the rest of the Wheatstone
circuit -- appear to fit well the experimental data. Obviously,
pure Josephson-coupled resistors are rather difficult to be
observed experimentally because the measured voltage is the sum of
contributions coming from the superconducting as well as the
semiconducting regions where the first ones are embedded. The case
of sample~1 is observed in $\sim 20\%$ of the measured lamellae.

\begin{figure}[t]   
\centering

\subfigure[]{\includegraphics[width=6cm]{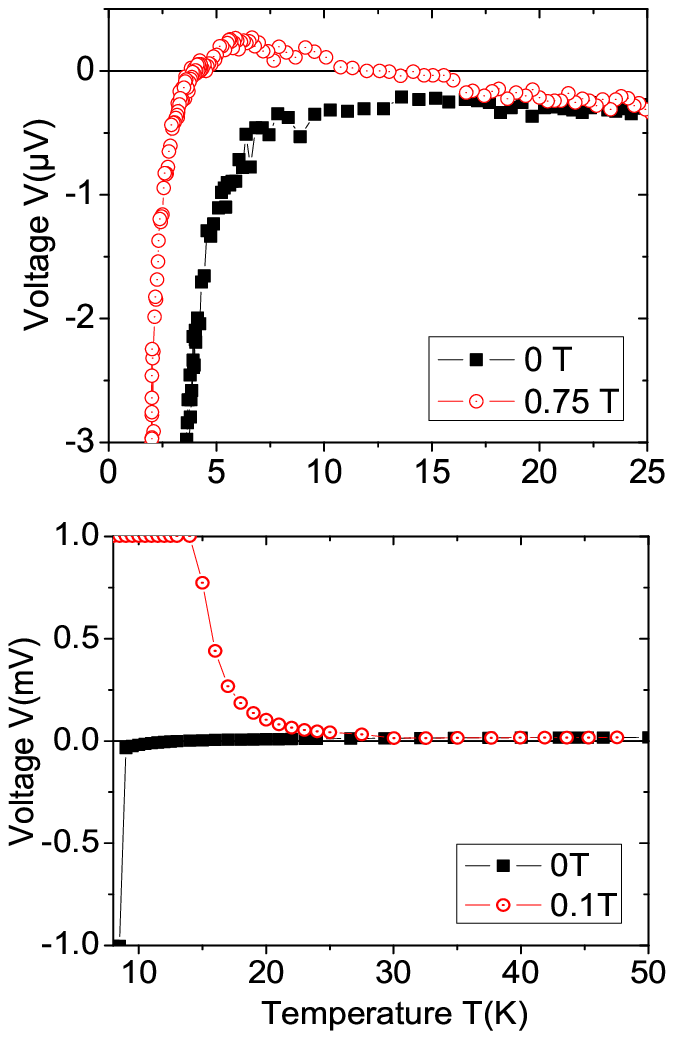}} 
\subfigure[]{\includegraphics[width=6cm]{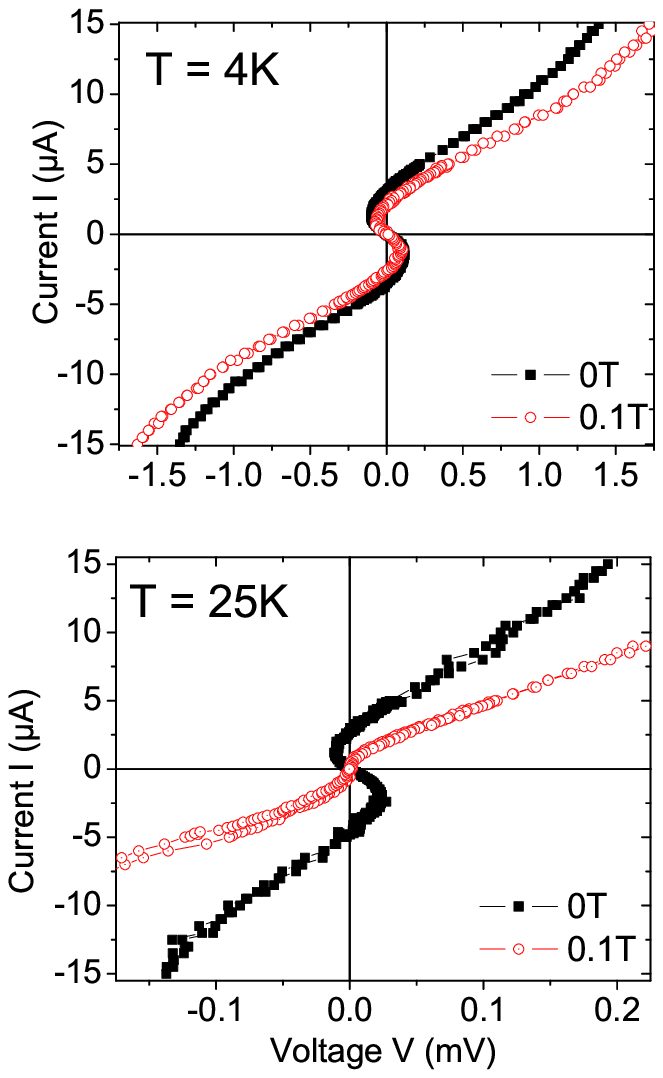}} 

\caption{(a) Voltage dependence on temperature for two different
lamellae. Upper picture: The transition denoted by the sharp
decrease in the voltage below a certain temperature (black
squares) is shifted to lower values when 0.75~T is applied
perpendicular to the planes (red dots). Bottom picture: Similar to
the other picture but for a different lamella.  The zero field
(black squares) transition is completely suppressed after applying
0.1~T (red dots). (b) Current-voltage characteristic curves for a
different lamella, with (red dots) and without field (black
squares) at 4~K (upper graph) and 25~K (bottom graph).}
\label{field}
\end{figure}

As shown in Section~\ref{ahloops} the role of the magnetic field
in the granular superconductivity behavior is definitively of huge
relevance. When superconducting patches are present between
semiconducting normal regions (we assume that they run mostly
parallel to the graphene layers) a remarkable effect must appear
when a magnetic field is applied in the c-axis direction and less
influence should be observed if it is applied parallel to the
graphene planes. The results in Fig.~\ref{field} correspond to the
situation where the field is applied parallel to the c-axis of the
lamellae. The in-plane case is not shown here because no effect
has been observed, as expected. Figure~\ref{field}(a) shows the
temperature behavior of the measured voltage with and without
field. The shift to lower temperatures of the transition after
applying 0.75~T is clearly shown for one of the samples. In the
case of other sample a 0.1~T field radically changes the
previously observed drop in voltage and a semiconducting-like
behavior appears. This behavior resembles the MIT found already in
HOPG and discussed in Section~\ref{bulkgra}. Figure~\ref{field}(b)
shows the CVC at 4~K and 25~K with and without field.  One can
recognize that the effect of the field is to reduce the
superconducting effect, i.e. the CVC tends to a linear, positive
voltage behavior.

\begin{figure}[t] 
\centering
\includegraphics[width=7.5cm]{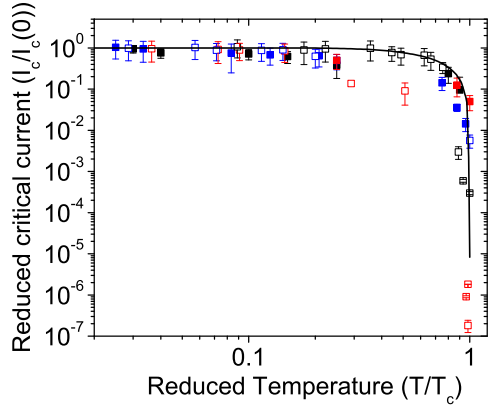}
\caption{Normalized Josephson critical current $I_c/I_c(0)$ as a
function of the normalized temperature $T/T_c$ obtained for
different samples. The values obtained for $I_c(0)$ range from 0.9
to $7~\mu$A and for $T_c \sim 23~$k to 175~K for different
samples. The solid line is the theoretical curve taken from the
work of \cite{hag10} without free parameters.}\label{ic}
\end{figure}

\cite{hee07} showed that Cooper pairs can flow in a single
graphene layer relatively long distances between two
superconducting contacts. The authors showed also the existence of
Josephson currents. These results support the main idea proposed
in this Section concerning the weakly Josephson coupled
superconducting regions through a graphene layer. In addition to
the work of \cite{hee07}, \cite{hag10} developed a theory for the
Josephson critical current in ballistic
Superconductor-Graphene-Superconductor structures. We use their
model to calculate the reduced temperature dependent critical
Josephson current. In Fig.~\ref{ic} the reader can see the solid
line obtained from the model proposed and the experimental data
taken from the experimental work described in this Section. The
model fits well the data. In the near future further
investigations concerning the effect of the magnetic field in
these lamellae are needed.

\section{Conclusion}\label{con}

The main messages of the work described in this chapter are: (1)
Several of the transport properties measured in large graphite
samples are not intrinsic of graphite but are influenced in a
large extent by interfaces and other defects and impurities.
Taking into account that one defect or impurity in a graphene
layer inside graphite can contribute with one carrier, it is clear
that the carrier densities  are  not intrinsic but are due to
defects and to a large extent probably to highly conducting
interfaces, internal as well as at the sample boundaries. We note
also that, as the large experimental evidence in semiconductors
indicates, even the quantum Hall effect observed in graphite is
actually not intrinsic of ideal graphite but can be attributed to
the highly conducting quasi-two dimensional electron systems at
the interfaces. Clearly, the details of this effect will depend on
the sample dimensions and quality, providing us a way to
understand experimental discrepancies found in literature. Our
work indicates that the intrinsic mobility of the carriers in the
graphene layers inside graphite are extraordinarily large $\mu
> 10^6~$cm$^2$/Vs and these carriers show ballistic transport in
the micrometer range even at room temperature.

(2) The first observation of superconductivity in doped graphite
goes back to 1965 when it was observed in the potassium graphite
intercalated compound C8K \citep{han65}. A considerable amount of
studies has reported this phenomenon in intercalated graphite
compounds or doped graphite \citep{wel05,eme05,kop04}, however the
superconducting properties of pure graphite are still under
discussion. In this chapter we have given a possible clarification
of this topic. By reducing the dimensions of the samples, the
intrinsic properties of graphite have been deeper investigated.
The overall idea coming out from all the experimental results can
be resumed as follows: bulk graphite samples cannot be considered
as a uniform electronic system, but it must be considered as a
semiconducting matrix with metallic as well as superconducting
domains in it. We note that already \cite{lu06} realized the
coexistence of insulating and conducting behaviors in bulk
graphite surfaces. From this newly acquired knowledge the observed
behaviors appear compatible with the existence of non-percolative
partially highly conducting, partially superconducting regions
coupled by Josephson-coupling through graphene planes. The
interfaces observed by TEM may have enough carrier density to
trigger quasi-two dimensional superconductivity.

Theoretical work that deals with superconductivity in graphite as
well as in graphene has been published in recent years. For
example, $p$-type superconductivity has been predicted to occur in
inhomogeneous regions of the graphite structure \citep{gon01} or
$d-$wave high-$T_c$ superconductivity based on resonance valence
bonds \citep{doni07}. Following a BCS approach in two dimensions
 critical temperatures $T_c \sim 60~$K have been
obtained if the density of conduction electrons per graphene plane
increases to $n \sim 10^{14}~$cm$^{-2}$, a density that might be
induced by defects and/or hydrogen ad-atoms \citep{garbcs09}.
Further predictions for superconductivity in graphene support the
premise that $ n > 10^{13}~$cm$^{-2}$ in order to reach $T_c
> 1~$K \citep{uch07,kop08}.  The interfaces observed by TEM might then be the regions
where enough carrier density exists to trigger quasi-two
dimensional superconductivity. In contrast to the basically 3D
superconductivity in intercalated graphitic compounds
\citep{csa05} we expect that superconductivity at quasi-2D
graphite interfaces as well as at doped surfaces \citep{han10} may
exist at much higher temperatures, partially because of the role
of the high-energy phonons in the 2D graphite structure itself
\citep{garbcs09}. Room temperature superconductivity with a $d+id$
pairing symmetry has been predicted to occur in doped graphene
with a carrier concentration $n \gtrsim 10^{14}~$cm$^{-2}$
\citep{pat10}. We note that if we take into account the density of
interfaces then a measured carrier density in bulk graphite
samples of $5 \times 10^{12}~$cm$^{-2}$ would mean an effective
density $ \sim 2 \times 10^{14}~$cm$^{-2}$ at the interfaces. And
last but not least we refer to a recent theoretical work from
\cite{kop11} where the authors emphasize that a topological
protected flat band in semimetals may promote superconductivity at
very high temperatures.


\end{document}